# Fabrication of photonic resonators in bulk 4H-SiC

Otto Cranwell Schaeper[1], Johannes E. Fröch[1,*], Sejeong Kim[2], Zhao Mu[3], Milos Toth[1,4], Weibo Gao[3,5], Igor Aharonovich[1,4]

[1]School of Mathematical and Physical Sciences, Faculty of Science, University of Technology Sydney, Ultimo, New South Wales 2007, Australia

[2]Department of Electrical and Electronic Engineering, University of Melbourne, VIC, 3010, Australia

[3]Division of Physics and Applied Physics, School of Physical and Mathematical Sciences, Nanyang Technological University, Singapore, 637371 Singapore

[4]ARC Centre of Excellence for Transformative Meta-Optical Systems (TMOS), University of Technology Sydney, Ultimo, New South Wales 2007, Australia

[5]The Photonics Institute and Centre for Disruptive Photonic Technologies, Nanyang Technological University, Singapore, 637371 Singapore

Corresponding Author: johannes.froech@uts.edu.au

**ABSTRACT**

The design and engineering of photonic architectures, suitable to enhance, collect and guide light on chip is needed for applications in quantum photonics and quantum optomechanics. In this work we apply a Faraday cage based oblique angle etch method to fabricate various functional photonic devices from 4H Silicon Carbide - a material that has attracted attention in recent years, due to its potential in optomechanics, nonlinear optics and quantum information. We detail the processing conditions and thoroughly address the geometrical and optical characteristics of the fabricated devices. Employing photoluminescence measurements we demonstrate high quality factors for suspended microring resonators of up to 3500 in the visible range. Such devices will be applicable in the future to augment the properties of SiC in integrated on chip quantum photonics.

**INTRODUCTION**

Silicon Carbide (SiC) has emerged in the recent decade as a distinguished front runner for nanophotonic and quantum photonic applications for several reasons.[1-4] First, epitaxial growth of high quality SiC on a wafer scale is industrially established, yielding integrated SiC photonics an elevated potential of scalability. Second, optically active defects in the material are known in the visible and the infrared range, thus overcoming a stringent criteria for long distance quantum protocols that require low transmission losses through optical fibers.[5] Third, a range of defects in this material are known to possess manipulatable spin properties, which is crucial towards the realization of optically addressable qubits.[6-17] Recent progress regarding the characterization of these defects have also shown exceptionally long coherence times, boosting their attractiveness and applicability as quantum memory.[18-21] Exceeding these major advantages, the material also has an exceptionally high refractive index of ~ 2.6 in the visible and IR, a piezoelectric response, strong higher order non-linearities,[22, 23] and can be reliably doped into p and n type forms. This expansive list of material properties gives SiC an almost unrivaled status as a material platform to realize nano photonic integrated circuits with potential applications in quantum technologies.

Yet, to fully leverage the potential of SiC quantum emitters, reliable fabrication approaches are essential for the engineering of photonic architectures and to augment their qubit properties.[1, 24-27] Accordingly, various fabrication methods and protocols have emerged over the recent years. Initially such schemes included the fabrication of devices from epitaxial SiC thin films grown on Si, which could be released by a standard wet chemical etch.[28-32] However, this is only applicable to the 3C polytype and moreover suffers from growth defects at the Si-SiC interface. As an alternative differently doped epitaxial layers of 4H-SiC were utilized, where the top layer could be released by a photoelectrochemical assisted etch.[33, 34] Yet, this scheme can also be seen as disadvantageous because of the presence of dopants which can deteriorate the spin properties of the qubit. Alternatively, the smart cut method was frequently demonstrated to fabricate SiC thin films of any polytype on an insulator platform.[35-38] However, in this method high dose ion implantation is used, introducing substantial damage, which is as well detrimental to the qubit properties. To overcome this problem, a pathway for integration of high quality epitaxial SiC on insulator has recently been shown. In this case bulk SiC is wafer bonded to a $SiO_2$ wafer and subsequently the bottom SiC side is reduced to the desired device layer thickness, forming a thin film of high quality SiC on insulator.[27, 39, 40] With this method Q-factors up to $10^6$ have been recently demonstrated in the IR range.[41]

As an alternative to the aforementioned methods, fabrication by angled plasma etching was explored, demonstrating photonic resonators with Q-factors on the order of $10^4$.[42] However, the devices were fabricated for operation in the IR range at 1550 nm, where no defects in SiC have been identified so far.

In this work, we build on previous efforts and demonstrate the fabrication of various free standing photonic resonators in bulk 4H-SiC with resonances in the visible range. Specifically, we demonstrate suspended microring resonators with Q-factors up to 3500 in the visible range.

**RESULTS and DISCUSSION**

The angle etch technique was initially developed for the fabrication of suspended diamond nanostructures.[43, 44] Here, we deploy the same methodology using a Faraday cage *in situ* a reactive ion etching (RIE) to impose an oblique etch directionality of ions onto silicon carbide. In the following, the fabrication steps are detailed, as schematically shown in Figure 1. First, a substrate of 4H-SiC (Norstel) was cleaned in hot piranha (H2O2: H2SO4, 1:3), then sequentially rinsed with acetone, isopropyl alcohol, water, and finally dried under nitrogen flow. Chemical semi amplified positive e-beam resist (CSAR-62, AllResist) was spin coated onto the substrate at 5000 rpm and baked at 180°C for 3 minutes, which resulted in a resist thickness of ~ 500 nm (Figure 1(a)). The desired design was then patterned into the resist by electron beam lithography (EBL - Zeiss Supra and RAITH EBL system). After development, 10 nm of Ti, followed by 230 nm of Ni was sputtered onto the sample. Following resist lift off in a remover (AR-600-546, AllResist), the metal remained as a mask on the substrate for the subsequent RIE steps, as shown in Figure 1 (b). A conical Faraday cage, machined from Al, shown in Figure 1(c), was then used for the ensuing etching steps to achieve undercut structures. Here we emphasize the importance of the right material choice for compatibility with the etch chemistry to avoid corrosion of the cage, which results in contamination and micromasking on the sample.

For the etching step, the sample was then positioned inside the cage in a central position as depicted in the inset. During the etching steps (Figure 1(d)), the sample was first exposed to a cleaning etch (5 min, 500 ICP, 15 RF, 10 sccm Ar, and 30 sccm $O_2$, 20mTorr). This pre-cleaning step is crucial to provide a clean surface by removing contaminants and thus avoiding uneven etch depths or micromasking, which would be detrimental to the device functionality. The sample was then etched (500 ICP, 15 RF, 3 sccm Ar, 3 sccm $O_2$, and 15 sccm of $SF_6$, 10mTorr) in increments of 4 minutes, for a total of 16 minutes, where the cage was rotated by 90° in subsequent steps. Incremental etching steps and cage rotation are required to average out irregularities in the cage geometry, as seen in Figure 1(c). In a final step (Figure 1(d)) the remaining metal mask was then removed by placing the sample in aqua regia (HCl:HNO3, 1:1).

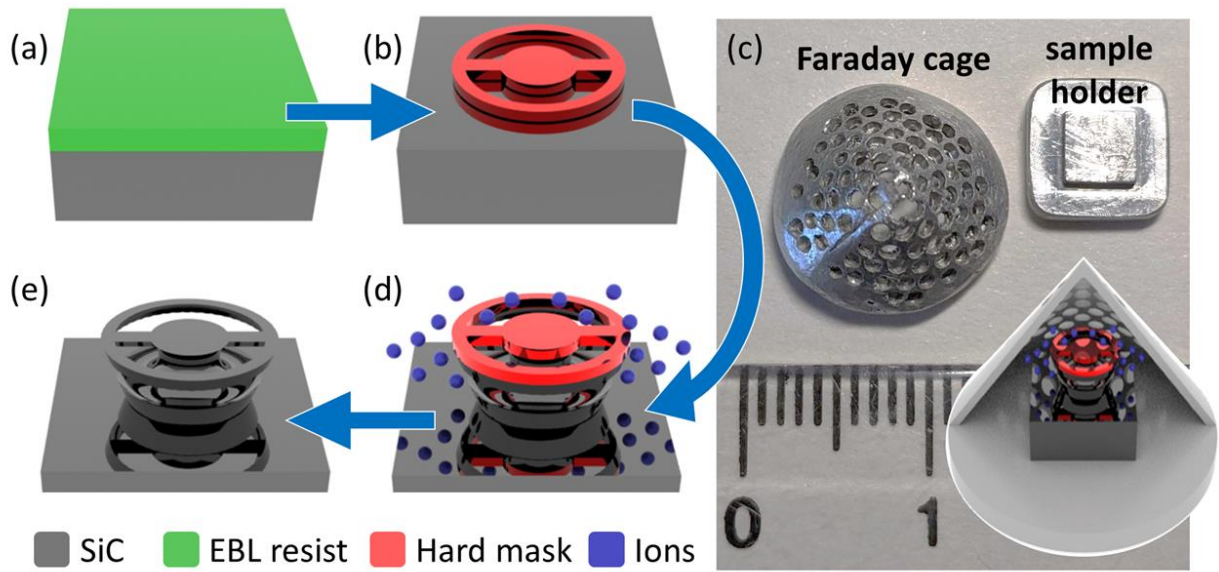

*Figure 1.* *Schematic of SiC photonic structure fabrication by angle etch RIE. (a) A 4H-SiC substrate (grey) is coated with EBL resist (green). (b) After metal (Ti and Ni) deposition and lift-off a hard mask (red) with the intended design remains on the sample. The sample is placed on the*

*sample holder inside a conical shaped cage as shown in a photo in (c). The inset shows the placement of the sample inside the Faraday cage. The ruler below the cage has a cm scale. (d) The sample is then etched with ions at an oblique angle. (e) After removing the metal mask using a wet chemical etch, the sample is finished.*

The versatility of this fabrication method and its applicability to bulk 4H-SiC is proven in the following by demonstrating a range of common photonic components used for on-chip integration. More specifically, we showcase devices based on a nanobeam geometry and ring resonators, achieving high device Quality-factors (Q-factor) in the visible range. As a representative base for nanobeam type devices, a conventional waveguide with notch coupler at the respective ends, is shown in Figure 2(a) in a top view. Such a waveguide is the essential component for on-chip photon routing. We note that such structures are typically fabricated with a prism shaped Faraday cage instead of a conical shaped Faraday cage as shown in Figure 1. The fabricated structures are indeed suspended and mechanically stable and do not bend after fabrication, as shown for a functionalized nanobeam in a tilted view (Figure 2(b)). Specifically for this device a refractive index variation is introduced in the waveguide, which enables the deliberate tailoring of the photonic bands in order to achieve either a spectral filtering element or a photonic crystal cavity (PCC). While this is commonly achieved by introducing air holes in the backbone of the structure, we implemented a beam width modulation, as shown in Figure 2(b), forming an alligator photonic crystal. Here broad regions are alternated periodically with thin regions, introducing dielectric and air like refractive index changes. Towards the center of the beam the periodicity is tapered, which is apparent in a magnified view of the device in Figure 2(c), depicting the cavity region and mirror region of such an alligator nanobeam cavity. As a guide to the eye the respective spacing of the photonic crystal, the tapering to form the cavity, and the central symmetry axis are indicated, by blue, red, and purple lines, respectively. The lattice modulation confines certain modes inside the nanobeam center, as shown by finite difference time domain (FDTD) simulations (Lumerical) in Figure 2(e) for the normalized E-field intensity. Therefore such a structure can act as a nanoscale photonic resonator with a high field intensity centered at the middle of the device, where a defect could be introduced for example by Focused Ion Beam implantation.[7, 45] We iterate that by maintaining the nanobeam shape without bending or the introduction of stresses and the capability to functionalize the structure, this fabrication approach will be appealing to SiC optomechanics in the future, as demonstrated before for diamond.[46]

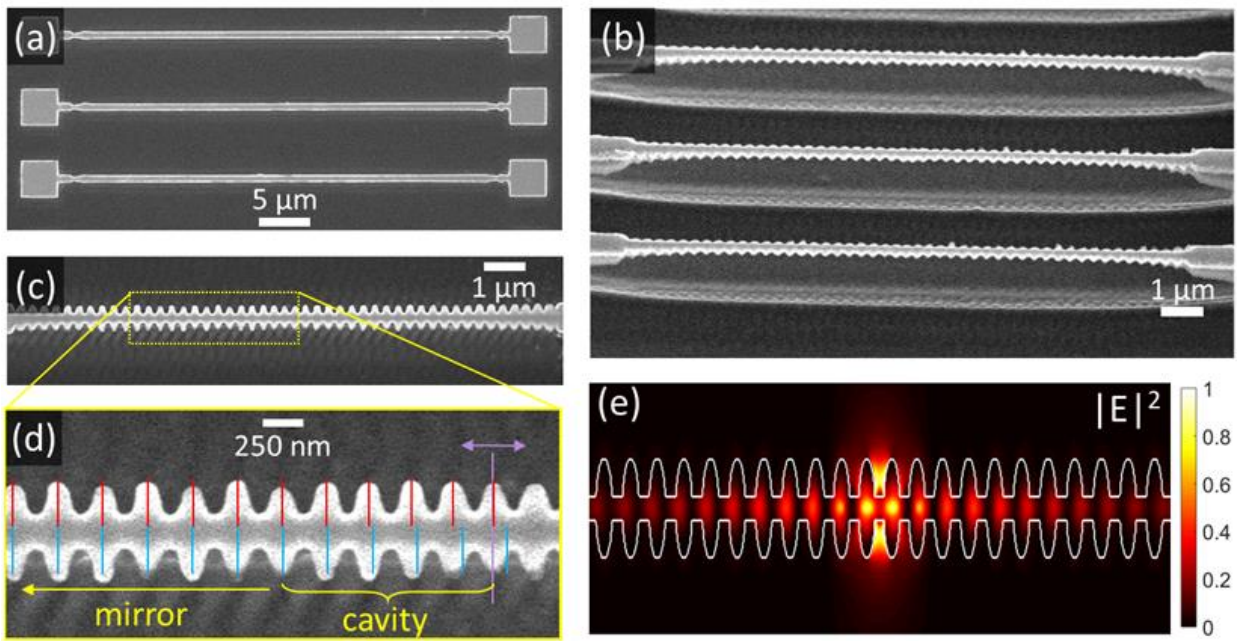

*Figure 2. SiC photonic nanobeam type structures. (a) Top view SEM image of an array of suspended nanobeam waveguides. (b) Tilted view of nanobeam devices. (c) Top view of an alligator shaped photonic nanobeam structure. (d) Magnified view of the device center, indicating the regions of the cavity and the photonic mirror. The periodicity of the photonic crystal is indicated by blue lines, while the actual distance between bars is indicated by red lines, depicting*

*well the tapering of the lattice towards the center. The central symmetry axis is indicated in purple. (e) FDTD simulation of the alligator PCC, shown in (c) and (d), depicting the normalized E-field intensity.*

Next, we move to demonstrate suspended microring resonators that are traditionally challenging to fabricate. Figure 3a shows an array of devices with different diameters, utilizing a design where the ring is connected by two nanoscale tethers to a central post, which holds the structure above the sample surface and thus physically isolating it from the substrate. This maximizes the refractive index contrast and facilitates high quality whispering gallery modes (WGM). A top view (Figure 3(b)) of a large microring (7.7 µm diameter) directly shows the significant size difference between a support tether of 180 nm width and the ring of 660 nm width. Because the tether does impart an optical loss on the WGM, it was crucial to minimize its size during fabrication, while maintaining a high yield of working devices. We note that, despite the successful suspension of all ring structures, different ring sizes may require different ideal etching times. Particularly, as can be seen in the same SEM image in Figure 3(a), the volume beneath smaller rings with the smallest diameter of 3 µm, appear with more material underneath. This is related to an RIE etch lag effect, for which smaller features (such as a smaller opening in the ring center) are etched at a lower rate, due to less accessibility of the ions to this region and a slower removal rate of the volatile etch products. Nevertheless, the applied method yields triangular shaped cross sections as shown in Figure 3(c), showing that the device geometry is dictated by the utilized Faraday cage.

FDTD analysis of the structures revealed and underlined two favorable properties of the fabricated devices. First, as shown in Figure 3(d) for the top view of the $|E|^2$ field intensity (log-scale) of the fundamental TE like mode, the WGM is well confined within the ring, while the tether only imparts a small amount of loss on it, evident by light coupling into the tether. Second, as shown in the cross sectional view for the fundamental TE and TM like modes, in Figure 3(e) and (f), respectively, one can see that the high field intensity is well centered around the middle of the triangle. This is important, because integrated spin qubits require placement as far as possible to any surface in order to avoid (or at least minimize) the detrimental influence of surface charge traps, which act disturbingly and reduce their coherence time.

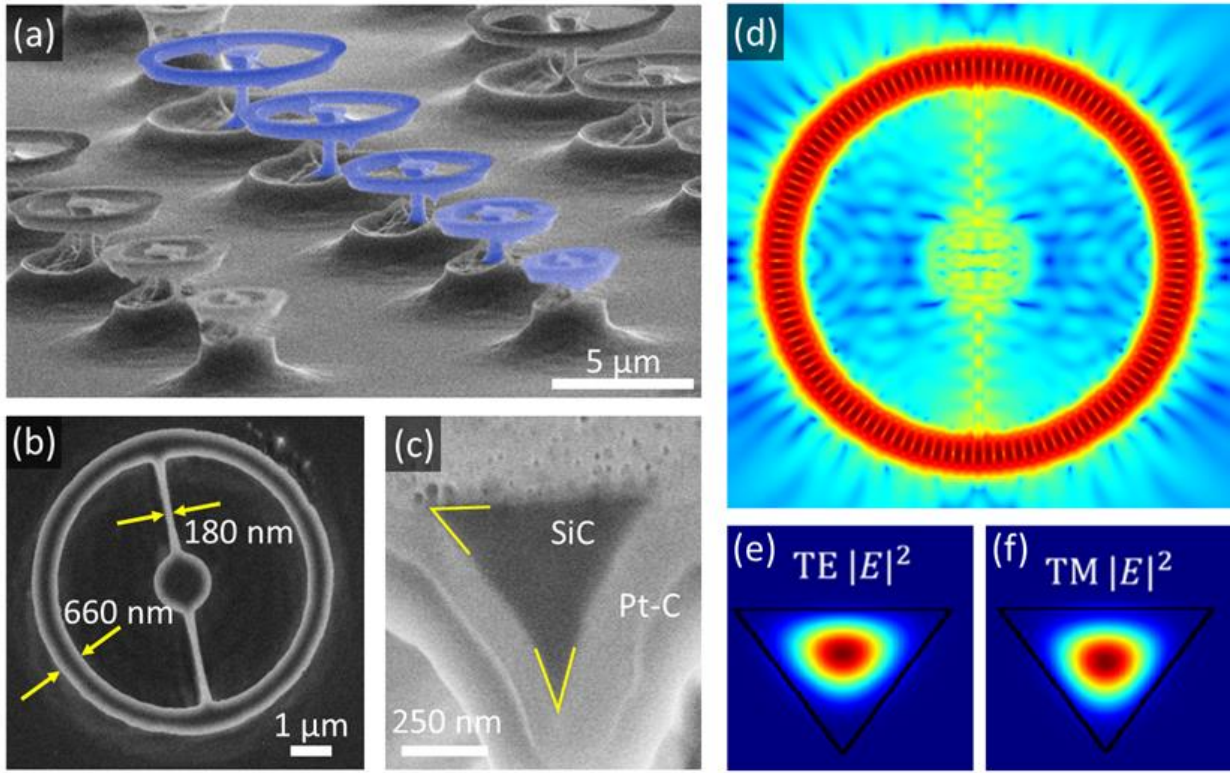

*Figure 3. Suspended ring resonators in 4H-SiC. (a) Tilted SEM image of an array of suspended microrings. (b) Top view of a large ring with 7.7 μm diameter, ring width of 660 nm and tether width of 180 nm. (c) Cross section of a ring showing the triangular shape of the microring. (d) Top view of a FDTD simulation for the fundamental TE mode shown on a log scale to demonstrate the low loss in the tether. Cross sectional view of the fundamental TE and TM like modes are shown in (e) and (f), respectively.*

In the following the device functionalities of microrings, as presented in Figure 2 were studied in a lab-built confocal photoluminescence (PL) setup. The background PL of the material was excited with a 1 mW, 532 nm laser through a 0.9 NA objective, and collected through the same objective. The emission was either guided to an APD or a spectrometer for analysis. For light that coupled to a resonator, distinguishable modes, superimposed on the background PL, could then be observed, as shown in Figure 4(a). Specifically, this spectrum was collected from the tether link, shown as a blue marked spot in the PL map (inset). Here it is apparent that the tether conveniently acts in addition as a local scatterer, which outcouples the light from the ring resonator. We note that we also observed an increased intensity at the central post. This is likely related to the support post being undercut, and acting itself as a resonator and allowing weakly confined WGM at the rim of the post. A high resolution spectrum for a device with a diameter of ~ 6 µm is shown in Figure 4(b). Here a family of modes (A - D) is apparent by equally spaced narrow resonances with a mode spacing of ~ 10 nm. We assume that the modes with the narrowest resonance correspond to either the fundamental TE or TM resonances, because they are confined at the center of the device, and are therefore less affected by surface roughness. In contrast higher order modes typically extend closer to the surface and thus are more prone to scattering from surface roughness, which results in a lower Q-factor. These resonances were further analyzed by multi-Lorentzian fits, which yielded Q-factors of ~ 1240, 2740, 2530, and 2440 at wavelengths of 708 nm, 719 nm, 729 nm, and 740 nm, respectively. The appearance of a neighboring narrow resonance at 730 nm with a Q-factor of 2000 is assumed to be of the vice versa mode family. Regardless, we emphasize that these devices are the first resonators from bulk 4H-SiC by angle plasma etching in the visible range, operating close to the emission of some relevant spin qubits, such as the $V^-_{Si}$ at 862-917 nm.

By further etch optimization a significant increase in Q-factor can be expected, making these devices even more suitable for emitter coupling.

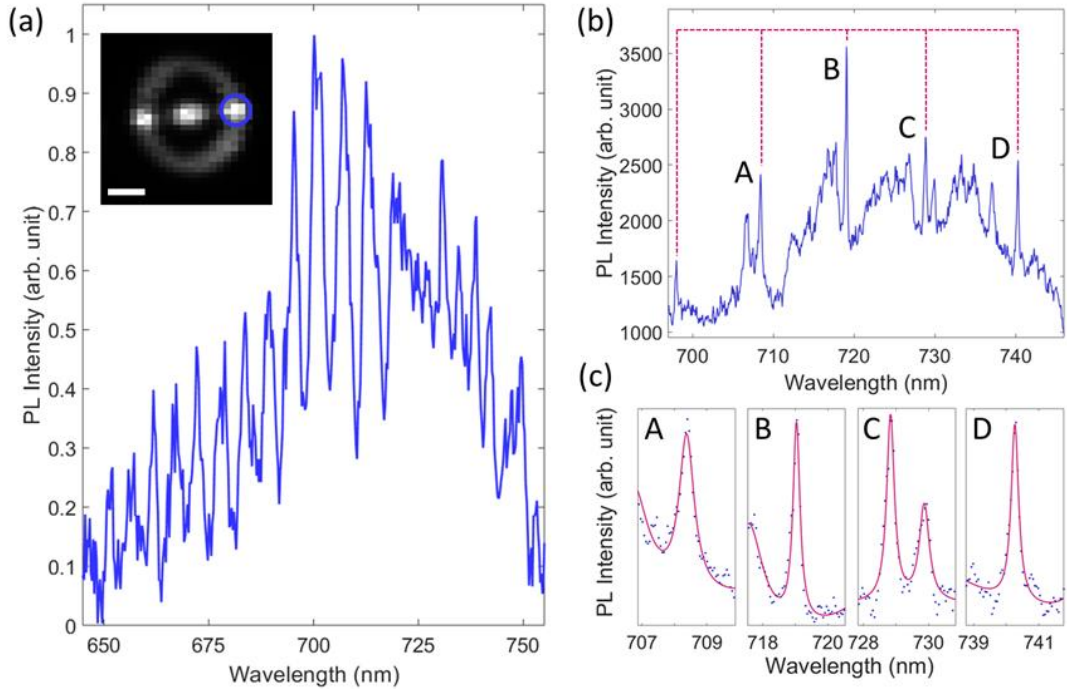

*Figure 4. PL analysis of individual devices. (a) Spectrum of a suspended Microring excited at the tether-ring link, as shown in the PL map inset. The scale bar in the inset corresponds to 2.5 µm. (b) High resolution spectrum of a microring showing recurring peaks with a spacing of ~ 10 nm. (c) Peaks of a recurring mode with the same periodicity are fitted by multiple Lorentzians.*

Considering the reproducibility of functional devices we now turn to a comparison of different microrings and their average performance. First, we compared microrings with a diameter of 3.7 µm (small), 4.7 µm, 5.7 µm (medium), 6.8 µm, and 7.7 µm (large). Representative spectra of a small, medium, and large ring are shown in Figure 5(a), in green (top), red (middle), and blue (bottom), respectively. To emphasize the periodicity, recurring peaks are highlighted. Naturally, for small devices, we observe clearly spaced peaks with a mode spacing on the order of $\sim$ 20 nm. Particularly, we clearly identified the fundamental modes, which appear pairwise within the shown spectral range. Whereas for medium and large rings, a smaller spacing of $\sim$ 10 nm and $\sim$ 6 nm, are observed, respectively. As summarized in Figure 5(b) on average, we observed Q-factors consistently above $10^3$ in all devices with medium rings displaying the highest Q-factor of 3500 and an average of $\sim$ 2800 ± 700 for medium rings. On the other hand as we decrease the ring radius the average Q-factor decreases to $\sim$ 2500 ± 300 for 4.7 µm and $\sim$ 1700 ± 500 for 3.7 µm. This decrease is commonly observed for WGM resonators and can be attributed to increased bending losses as the curvature of the ring increases. In addition, for smaller devices we observed an increased tether width, which induces higher optical losses due to scattering. Besides the decrease in Q-factor for smaller devices, we also observed a trend of decreasing Q-factor for devices with larger diameter. For which we determined average Q-factors of $\sim$ 2500 ± 300 and $\sim$ 1900 ± 400 for rings with 6.8 µm and 7.7 µm, respectively. Although, at first glance this trend seems counterintuitive we attribute it to the nature of the fabrication process. Specifically, as discussed before in Figure 1, microrings with larger sizes are etched at a higher rate, due to larger outlines and higher accessibility for ions and therefore higher removal rates. This in turn implies that for larger structures, where sidewalls are already undercut, stray ions or backscattered ions can induce further etching, which introduce a higher degree of surface damage in structures that are undercut at a faster rate. This exemplifies well how the processing parameters and geometries for the Faraday cage assisted angle fabrication have to be specifically tailored to achieve optimum results.

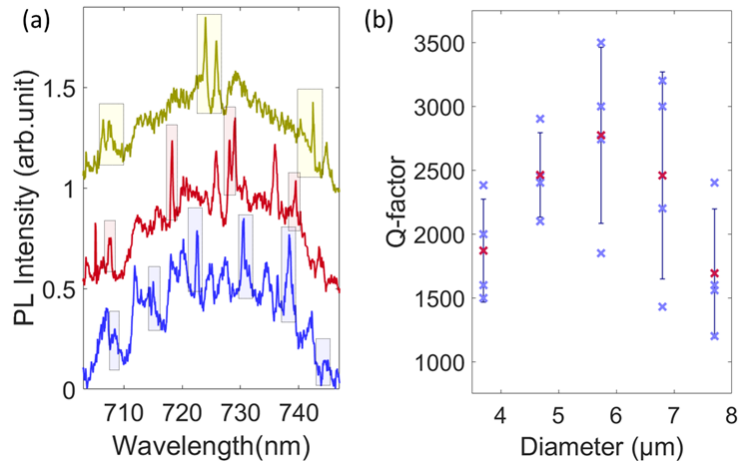

*Figure 5.* *Size dependence. (a) Spectra of rings with diameter of 3.7 μm, 5.7 μm, and 5.7 μm are shown in green, red, and blue, respectively (in that order from top to bottom). Recurring modes are highlighted to emphasize their periodicity. (b) Measurement of the highest Q-factors from various microrings with different size, indicated as blue crosses. The average Q-factors for each particular size are shown as red crosses. The error bars correspond to the standard deviation of the set.*

**CONCLUSION**

In summary, we demonstrated the nanofabrication of suspended photonic structures in high quality bulk 4H-SiC utilizing an oblique plasma etch method. Specifically, a Faraday cage setup is used to impose a custom anisotropic ion etch directionality in a non-trivial manner, which enables the fabrication of prototypical photonic devices such as nanobeam cavities and microring resonators. We describe in detail the conditions of the fabrication process and their implications on the device geometry. From FDTD simulations, we determine the device characteristics and relate the implications of their utilized structure on their functionality, showcasing optimized geometries. The device functionalities for microring resonators were determined in confocal PL measurements, demonstrating devices operating in the visible range and various sizes with consistent Q-factors above 1500 up to ~3500. With further improvement and scaling to the NIR range, we expect to improve the device functionality significantly, which will find applications in providing suitable photonic architectures to control and enhance spin qubits in SiC.

ACKNOWLEDGEMENTS

The authors thank Kevin Chen, Dirk Englund, Michael Walsh and Noel Wan for the initial assistance with the cage design. The authors thank the Australian Research Council (CE200100010) for the financial support. The authors would like to thank the ANFF UTS hub for access to their facilities.

REFERENCES

1.  Lukin, D. M.; Guidry, M. A.; Vučković, J., Integrated Quantum Photonics with Silicon Carbide: Challenges and Prospects. *PRX Quantum* **2020,** *1*, 020102.
2.  Castelletto, S.; Boretti, A., Silicon Carbide Color Centers for Quantum Applications. *Journal of Physics: Photonics* **2020,** *2*, 022001.
3.  Son, N. T.; Anderson, C. P.; Bourassa, A.; Miao, K. C.; Babin, C.; Widmann, M.; Niethammer, M.; Ul Hassan, J.; Morioka, N.; Ivanov, I. G.; Kaiser, F.; Wrachtrup, J.; Awschalom, D. D., Developing Silicon Carbide for Quantum Spintronics. *Appl. Phys. Lett.* **2020,** *116*, 190501.
4.  Radulaski, M.; Widmann, M.; Niethammer, M.; Zhang, J. L.; Lee, S.-Y.; Rendler, T.; Lagoudakis, K. G.; Son, N. T.; Janzén, E.; Ohshima, T.; Wrachtrup, J.; Vučković, J., Scalable Quantum Photonics with Single Color Centers in Silicon Carbide. *Nano Lett.* **2017,** *17*, 1782-1786.
5.  Lohrmann, A.; Johnson, B. C.; McCallum, J. C.; Castelletto, S., A Review on Single Photon Sources in Silicon Carbide. *Reports on Progress in Physics* **2017,** *80*, 034502.
6.  Widmann, M.; Lee, S.-Y.; Rendler, T.; Son, N. T.; Fedder, H.; Paik, S.; Yang, L.-P.; Zhao, N.; Yang, S.; Booker, I.; Denisenko, A.; Jamali, M.; Momenzadeh, S. A.; Gerhardt, I.; Ohshima, T.; Gali, A.; Janzén, E.; Wrachtrup, J., Coherent Control of Single Spins in Silicon Carbide at Room Temperature. *Nature Materials* **2014,** *14*, 164.
7.  Mu, Z.; Zargaleh, S. A.; von Bardeleben, H. J.; Fröch, J. E.; Nonahal, M.; Cai, H.; Yang, X.; Yang, J.; Li, X.; Aharonovich, I.; Gao, W., Coherent Manipulation with Resonant Excitation and Single Emitter Creation of Nitrogen Vacancy Centers in 4h Silicon Carbide. *Nano Lett.* **2020,** *20*, 6142-6147.

8. Wang, J.-F.; Yan, F.-F.; Li, Q.; Liu, Z.-H.; Liu, H.; Guo, G.-P.; Guo, L.-P.; Zhou, X.; Cui, J.-M.; Wang, J.; Zhou, Z.-Q.; Xu, X.-Y.; Xu, J.-S.; Li, C.-F.; Guo, G.-C., Coherent Control of Nitrogen-Vacancy Center Spins in Silicon Carbide at Room Temperature. *Phys. Rev. Lett.* **2020,** *124*, 223601.
9. Diler, B.; Whiteley, S. J.; Anderson, C. P.; Wolfowicz, G.; Wesson, M. E.; Bielejec, E. S.; Joseph Heremans, F.; Awschalom, D. D., Coherent Control and High-Fidelity Readout of Chromium Ions in Commercial Silicon Carbide. *npj Quantum Information* **2020,** *6*, 11.
10. Nagy, R.; Niethammer, M.; Widmann, M.; Chen, Y.-C.; Udvarhelyi, P.; Bonato, C.; Hassan, J. U.; Karhu, R.; Ivanov, I. G.; Son, N. T.; Maze, J. R.; Ohshima, T.; Soykal, Ö. O.; Gali, Á.; Lee, S.-Y.; Kaiser, F.; Wrachtrup, J., High-Fidelity Spin and Optical Control of Single Silicon-Vacancy Centres in Silicon Carbide. *Nat. Commun.* **2019,** *10*, 1954.
11. Christle, D. J.; Klimov, P. V.; de las Casas, C. F.; Szász, K.; Ivády, V.; Jokubavicius, V.; Ul Hassan, J.; Syväjärvi, M.; Koehl, W. F.; Ohshima, T.; Son, N. T.; Janzén, E.; Gali, Á.; Awschalom, D. D., Isolated Spin Qubits in Sic with a High-Fidelity Infrared Spin-to-Photon Interface. *Phys. Rev. X* **2017,** *7*, 021046.
12. Koehl, W. F.; Diler, B.; Whiteley, S. J.; Bourassa, A.; Son, N. T.; Janzén, E.; Awschalom, D. D., Resonant Optical Spectroscopy and Coherent Control of $\mathrm{C}{\mathrm{R}}^{4+}$ Spin Ensembles in Sic and Gan. *Physical Review B* **2017,** *95*, 035207.
13. Soltamov, V. A.; Yavkin, B. V.; Tolmachev, D. O.; Babunts, R. A.; Badalyan, A. G.; Davydov, V. Y.; Mokhov, E. N.; Proskuryakov, I. I.; Orlinskii, S. B.; Baranov, P. G., Optically Addressable Silicon Vacancy-Related Spin Centers in Rhombic Silicon Carbide with High Breakdown Characteristics and Endor Evidence of Their Structure. *Phys. Rev. Lett.* **2015,** *115*, 247602.
14. Soltamov, V. A.; Kasper, C.; Poshakinskiy, A. V.; Anisimov, A. N.; Mokhov, E. N.; Sperlich, A.; Tarasenko, S. A.; Baranov, P. G.; Astakhov, G. V.; Dyakonov, V., Excitation and Coherent Control of Spin Qudit Modes in Silicon Carbide at Room Temperature. *Nat. Commun.* **2019,** *10*, 1678.
15. Fuchs, F.; Stender, B.; Trupke, M.; Simin, D.; Pflaum, J.; Dyakonov, V.; Astakhov, G. V., Engineering near-Infrared Single-Photon Emitters with Optically Active Spins in Ultrapure Silicon Carbide. *Nat. Commun.* **2015,** *6*, 7578.
16. Banks, H. B.; Soykal, Ö. O.; Myers-Ward, R. L.; Gaskill, D. K.; Reinecke, T. L.; Carter, S. G., Resonant Optical Spin Initialization and Readout of Single Silicon Vacancies in $4h$-$\mathrm{Si}\mathrm{C}$. *Physical Review Applied* **2019,** *11*, 024013.
17. Wolfowicz, G.; Anderson, C. P.; Diler, B.; Poluektov, O. G.; Heremans, F. J.; Awschalom, D. D., Vanadium Spin Qubits as Telecom Quantum Emitters in Silicon Carbide. *Science Advances* **2020,** *6*, eaaz1192.
18. Simin, D.; Kraus, H.; Sperlich, A.; Ohshima, T.; Astakhov, G. V.; Dyakonov, V., Locking of Electron Spin Coherence above 20 Ms in Natural Silicon Carbide. *Physical Review B* **2017,** *95*, 161201.
19. Miao, K. C.; Blanton, J. P.; Anderson, C. P.; Bourassa, A.; Crook, A. L.; Wolfowicz, G.; Abe, H.; Ohshima, T.; Awschalom, D. D., Universal Coherence Protection in a Solid-State Spin Qubit. *Science* **2020,** *369*, 1493.
20. Christle, D. J.; Falk, A. L.; Andrich, P.; Klimov, P. V.; Hassan, J. U.; Son, Nguyen T.; Janzén, E.; Ohshima, T.; Awschalom, D. D., Isolated Electron Spins in Silicon Carbide with Millisecond Coherence Times. *Nature Materials* **2015,** *14*, 160-163.

21. Anderson, C. P.; Bourassa, A.; Miao, K. C.; Wolfowicz, G.; Mintun, P. J.; Crook, A. L.; Abe, H.; Ul Hassan, J.; Son, N. T.; Ohshima, T.; Awschalom, D. D., Electrical and Optical Control of Single Spins Integrated in Scalable Semiconductor Devices. *Science* **2019**, *366*, 1225.
22. Sato, H.; Abe, M.; Shoji, I.; Suda, J.; Kondo, T., Accurate Measurements of Second-Order Nonlinear Optical Coefficients of 6h and 4h Silicon Carbide. *J. Opt. Soc. Am. B* **2009**, *26*, 1892-1896.
23. Lu, X.; Lee, J. Y.; Rogers, S.; Lin, Q., Optical Kerr Nonlinearity in a High-Q Silicon Carbide Microresonator. *Opt. Express* **2014**, *22*, 30826-30832.
24. Wang, Y.; Lin, Q.; Feng, P. X. L., Single-Crystal 3c-Sic-on-Insulator Platform for Integrated Quantum Photonics. *Opt. Express* **2021**, *29*, 1011-1022.
25. Majety, S.; Norman, V. A.; Li, L.; Bell, M.; Saha, P.; Radulaski, M., Quantum Photonics in Triangular-Cross-Section Nanodevices in Silicon Carbide. *arXiv e-prints* **2020**, arXiv:2012.02350.
26. Crook, A. L.; Anderson, C. P.; Miao, K. C.; Bourassa, A.; Lee, H.; Bayliss, S. L.; Bracher, D. O.; Zhang, X.; Abe, H.; Ohshima, T.; Hu, E. L.; Awschalom, D. D., Purcell Enhancement of a Single Silicon Carbide Color Center with Coherent Spin Control. *Nano Lett.* **2020**, *20*, 3427-3434.
27. Lukin, D. M.; Dory, C.; Guidry, M. A.; Yang, K. Y.; Mishra, S. D.; Trivedi, R.; Radulaski, M.; Sun, S.; Vercruysse, D.; Ahn, G. H.; Vučković, J., 4h-Silicon-Carbide-on-Insulator for Integrated Quantum and Nonlinear Photonics. *Nat. Photonics* **2019**.
28. Radulaski, M.; Babinec, T. M.; Buckley, S.; Rundquist, A.; Provine, J.; Alassaad, K.; Ferro, G.; Vučković, J., Photonic Crystal Cavities in Cubic (3c) Polytype Silicon Carbide Films. *Opt. Express* **2013**, *21*, 32623-32629.
29. Cardenas, J.; Zhang, M.; Phare, C. T.; Shah, S. Y.; Poitras, C. B.; Guha, B.; Lipson, M., High Q Sic Microresonators. *Opt. Express* **2013**, *21*, 16882-16887.
30. Lu, X.; Lee, J. Y.; Feng, P. X. L.; Lin, Q., High Q Silicon Carbide Microdisk Resonator. *Appl. Phys. Lett.* **2014**, *104*, 181103.
31. Radulaski, M.; Babinec, T. M.; Müller, K.; Lagoudakis, K. G.; Zhang, J. L.; Buckley, S.; Kelaita, Y. A.; Alassaad, K.; Ferro, G.; Vučković, J., Visible Photoluminescence from Cubic (3c) Silicon Carbide Microdisks Coupled to High Quality Whispering Gallery Modes. *ACS Photonics* **2015**, *2*, 14-19.
32. Chatzopoulos, I.; Martini, F.; Cernansky, R.; Politi, A., High-Q/V Photonic Crystal Cavities and Qed Analysis in 3c-Sic. *ACS Photonics* **2019**, *6*, 1826-1831.
33. Magyar, A. P.; Bracher, D.; Lee, J. C.; Aharonovich, I.; Hu, E. L., High Quality Sic Microdisk Resonators Fabricated from Monolithic Epilayer Wafers. *Appl. Phys. Lett.* **2014**, *104*, 051109.
34. Bracher, D. O.; Hu, E. L., Fabrication of High-Q Nanobeam Photonic Crystals in Epitaxially Grown 4h-Sic. *Nano Lett.* **2015**, *15*, 6202-6207.
35. Song, B. S.; Yamada, S.; Asano, T.; Noda, S., Demonstration of Two-Dimensional Photonic Crystals Based on Silicon Carbide. *Opt. Express* **2011**, *19*, 11084-11089.
36. Yi, A.; Zheng, Y.; Huang, H.; Lin, J.; Yan, Y.; You, T.; Huang, K.; Zhang, S.; Shen, C.; Zhou, M.; Huang, W.; Zhang, J.; Zhou, S.; Ou, H.; Ou, X., Wafer-Scale 4h-Silicon Carbide-on-Insulator (4h–Sicoi) Platform for Nonlinear Integrated Optical Devices. *Optical Materials* **2020**, *107*, 109990.
37. Cardenas, J.; Yu, M.; Okawachi, Y.; Poitras, C. B.; Lau, R. K. W.; Dutt, A.; Gaeta, A. L.; Lipson, M., Optical Nonlinearities in High-Confinement Silicon Carbide Waveguides. *Opt. Lett.* **2015**, *40*, 4138-4141.

38. Yamada, S.; Song, B.-S.; Jeon, S.; Upham, J.; Tanaka, Y.; Asano, T.; Noda, S., Second-Harmonic Generation in a Silicon-Carbide-Based Photonic Crystal Nanocavity. *Opt. Lett.* **2014,** *39*, 1768-1771.
39. Song, B.-S.; Asano, T.; Jeon, S.; Kim, H.; Chen, C.; Kang, D. D.; Noda, S., Ultrahigh-Q Photonic Crystal Nanocavities Based on 4h Silicon Carbide. *Optica* **2019,** *6*, 991-995.
40. Fan, T.; Wu, X.; Eftekhar, A. A.; Bosi, M.; Moradinejad, H.; Woods, E. V.; Adibi, A., High-Quality Integrated Microdisk Resonators in the Visible-to-near-Infrared Wavelength Range on a 3c-Silicon Carbide-on-Insulator Platform. *Opt. Lett.* **2020,** *45*, 153-156.
41. Guidry, M. A.; Yang, K. Y.; Lukin, D. M.; Markosyan, A.; Yang, J.; Fejer, M. M.; Vučković, J., Optical Parametric Oscillation in Silicon Carbide Nanophotonics. *Optica* **2020,** *7*, 1139-1142.
42. Song, B.-S.; Jeon, S.; Kim, H.; Kang, D. D.; Asano, T.; Noda, S., High-Q-Factor Nanobeam Photonic Crystal Cavities in Bulk Silicon Carbide. *Appl. Phys. Lett.* **2018,** *113*, 231106.
43. Burek, M. J.; Chu, Y.; Liddy, M. S. Z.; Patel, P.; Rochman, J.; Meesala, S.; Hong, W.; Quan, Q.; Lukin, M. D.; Lončar, M., High Quality-Factor Optical Nanocavities in Bulk Single-Crystal Diamond. *Nat. Commun.* **2014,** *5*, 5718.
44. Burek, M. J.; de Leon, N. P.; Shields, B. J.; Hausmann, B. J. M.; Chu, Y.; Quan, Q.; Zibrov, A. S.; Park, H.; Lukin, M. D.; Lončar, M., Free-Standing Mechanical and Photonic Nanostructures in Single-Crystal Diamond. *Nano Lett.* **2012**.
45. Wang, J.; Zhou, Y.; Zhang, X.; Liu, F.; Li, Y.; Li, K.; Liu, Z.; Wang, G.; Gao, W., Efficient Generation of an Array of Single Silicon-Vacancy Defects in Silicon Carbide. *Physical Review Applied* **2017,** *7*, 064021.
46. Burek, M. J.; Cohen, J. D.; Meenehan, S. M.; El-Sawah, N.; Chia, C.; Ruelle, T.; Meesala, S.; Rochman, J.; Atikian, H. A.; Markham, M.; Twitchen, D. J.; Lukin, M. D.; Painter, O.; Lončar, M., Diamond Optomechanical Crystals. *Optica* **2016,** *3*, 1404-1411.